\documentclass[10pt,twocolumn]{article}

\usepackage{graphicx}
\usepackage{amsmath}
\usepackage{url}
\usepackage{hyperref}
\usepackage{booktabs}
\usepackage{xcolor}

\usepackage{url}
\usepackage{hyperref}

\title{Need for Speed: A Comprehensive Benchmark of JPEG Decoders in Python\thanks{\url{https://github.com/ternaus/imread_benchmark}}}

\author{
    Vladimir Iglovikov\\
    Ternaus Inc\\
    \texttt{vladimir@ternaus.com}
}

\begin{document}

\maketitle

\begin{abstract}
Image loading represents a critical bottleneck in modern machine learning pipelines, particularly in computer vision tasks where JPEG remains the dominant format. This study presents a systematic performance analysis of nine popular Python JPEG decoding libraries on different computing architectures. We benchmark traditional image processing libraries (Pillow, OpenCV), machine learning frameworks (TensorFlow, PyTorch), and specialized decoders (jpeg4py, kornia-rs) on both ARM64 (Apple M4 Max) and x86\_64 (AMD Threadripper) platforms. Our findings reveal that modern implementations using libjpeg-turbo achieve up to 1.5x faster decoding speeds compared to traditional approaches. We provide evidence-based recommendations for choosing optimal JPEG decoders across different scenarios, from high-throughput training pipelines to real-time applications. This comprehensive analysis helps practitioners make informed decisions about image loading infrastructure, potentially reducing training times and improving system efficiency.
\end{abstract}

\section{Introduction}
JPEG remains the dominant image format in the digital world \cite{wallace1992jpeg}, striking an optimal balance between compression efficiency and image quality. Despite the emergence of newer formats like WebP and AVIF, JPEG's ubiquity persists across web platforms, digital cameras, and mobile devices. This widespread use makes JPEG decoding performance a critical consideration for many applications.

In the Python ecosystem, developers face a choice between multiple libraries capable of JPEG decoding. These range from general-purpose image processing libraries like Pillow and OpenCV \cite{opencv_library} to specialized components of machine learning frameworks such as TensorFlow and PyTorch \cite{pytorch2019}. Each library implements JPEG decoding differently, potentially leading to significant performance variations.

The speed of JPEG decoding is crucial across numerous real-world applications. In machine learning pipelines, image loading can become a significant bottleneck during training. Web services must efficiently process large volumes of user-uploaded images, while real-time applications require immediate image processing capabilities. Large-scale batch processing systems often handle millions of images, and content delivery networks need to optimize image serving for optimal performance.

Despite the importance of JPEG decoding performance, comprehensive and up-to-date benchmarks comparing Python libraries are scarce. Most existing comparisons focus on feature sets rather than performance metrics, quickly become outdated as libraries evolve, lack cross-platform performance analysis, and often miss the impact of underlying JPEG decoder implementations.

The complete benchmark implementation and all test results are available as open source at \url{https://github.com/ternaus/imread_benchmark}.

This study presents a systematic benchmark of popular Python libraries for JPEG decoding across several categories:
\begin{itemize}
    \item Traditional image processing libraries
    \begin{itemize}
        \item Pillow and its SIMD-optimized variant (Pillow-SIMD, Linux only)
        \item OpenCV
    \end{itemize}
    \item Scientific computing libraries
    \begin{itemize}
        \item scikit-image
        \item imageio
    \end{itemize}
    \item Machine learning framework components
    \begin{itemize}
        \item TensorFlow
        \item torchvision
        \item kornia-rs \cite{kornia2020} (Rust implementation)
    \end{itemize}
    \item Specialized JPEG decoders
    \begin{itemize}
        \item jpeg4py (direct libjpeg-turbo binding, Linux only)
    \end{itemize}
\end{itemize}

\section{Background and Methodology}

\subsection{JPEG Decoding Overview}
JPEG decoding involves several key steps: entropy decoding (Huffman), inverse quantization, and inverse Discrete Cosine Transform (IDCT) \cite{wallace1992jpeg}. The performance of these operations can be significantly impacted by the implementation approach and optimizations used .

A critical component in JPEG decoding is the underlying decoder library. Most Python libraries rely on either:
\begin{itemize}
    \item libjpeg: The original reference implementation \cite{wallace1992jpeg}
    \item libjpeg-turbo: A highly optimized fork that uses SIMD instructions
    \item Custom implementations: Some libraries implement their own decoders or use alternative approaches
\end{itemize}

\subsection{Benchmark Setup}
Our benchmark measures the end-to-end performance of loading and decoding JPEG images, including disk I/O. This approach reflects real-world usage where both file system operations and decoding affect overall performance.

\subsubsection{Test Environment}
\begin{itemize}
    \item ARM64: Apple M4 Max (macOS 14.3)
    \item x86\_64: AMD Threadripper 3970X (Ubuntu 22.04)
\end{itemize}

\subsubsection{Dataset}
\begin{itemize}
    \item 2000 images from ImageNet validation set \cite{imagenet}
    \item Average file size: 130.2 KB
    \item Resolution range: 90x90 - 3264x2792 pixels
\end{itemize}

\subsubsection{Methodology}
\begin{itemize}
    \item 20 runs per library
    \item Cold start for each run (clearing file system cache)
    \item Measuring total time for loading and decoding
    \item Statistical analysis including mean and standard deviation
\end{itemize}

\section{Results}

Our benchmark reveals significant performance differences between libraries, influenced by their implementation approaches and underlying JPEG decoders. Tests were conducted on two distinct platforms: Apple M4 Max (ARM64) and AMD Threadripper 3970X (x86\_64).

\subsection{Performance Comparison}

\begin{figure}[t]
    \centering
    \includegraphics[width=\columnwidth]{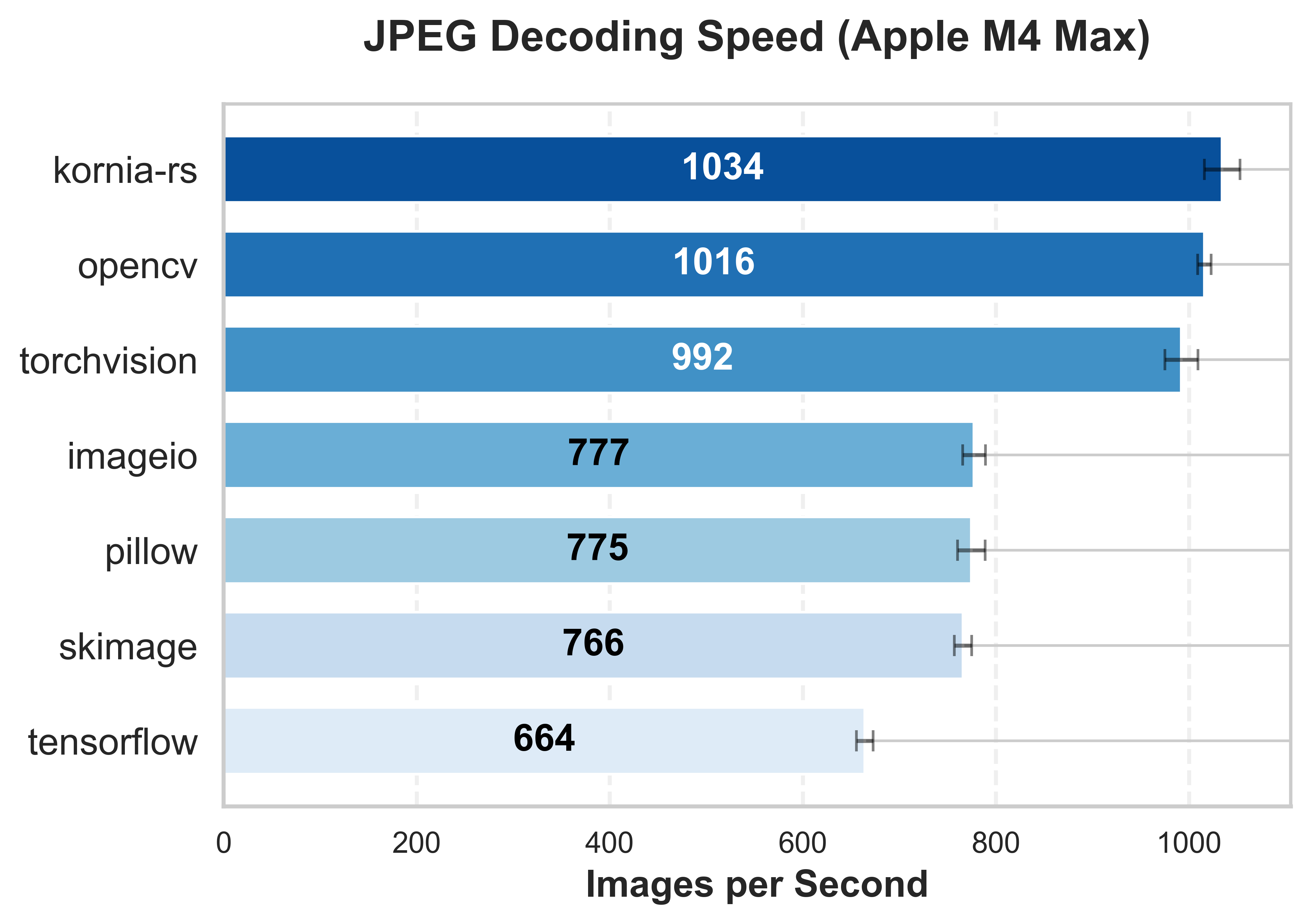}
    \caption{JPEG Decoding Performance on Apple M4 Max}
    \label{fig:perf_darwin}
\end{figure}

\begin{figure}[t]
    \centering
    \includegraphics[width=\columnwidth]{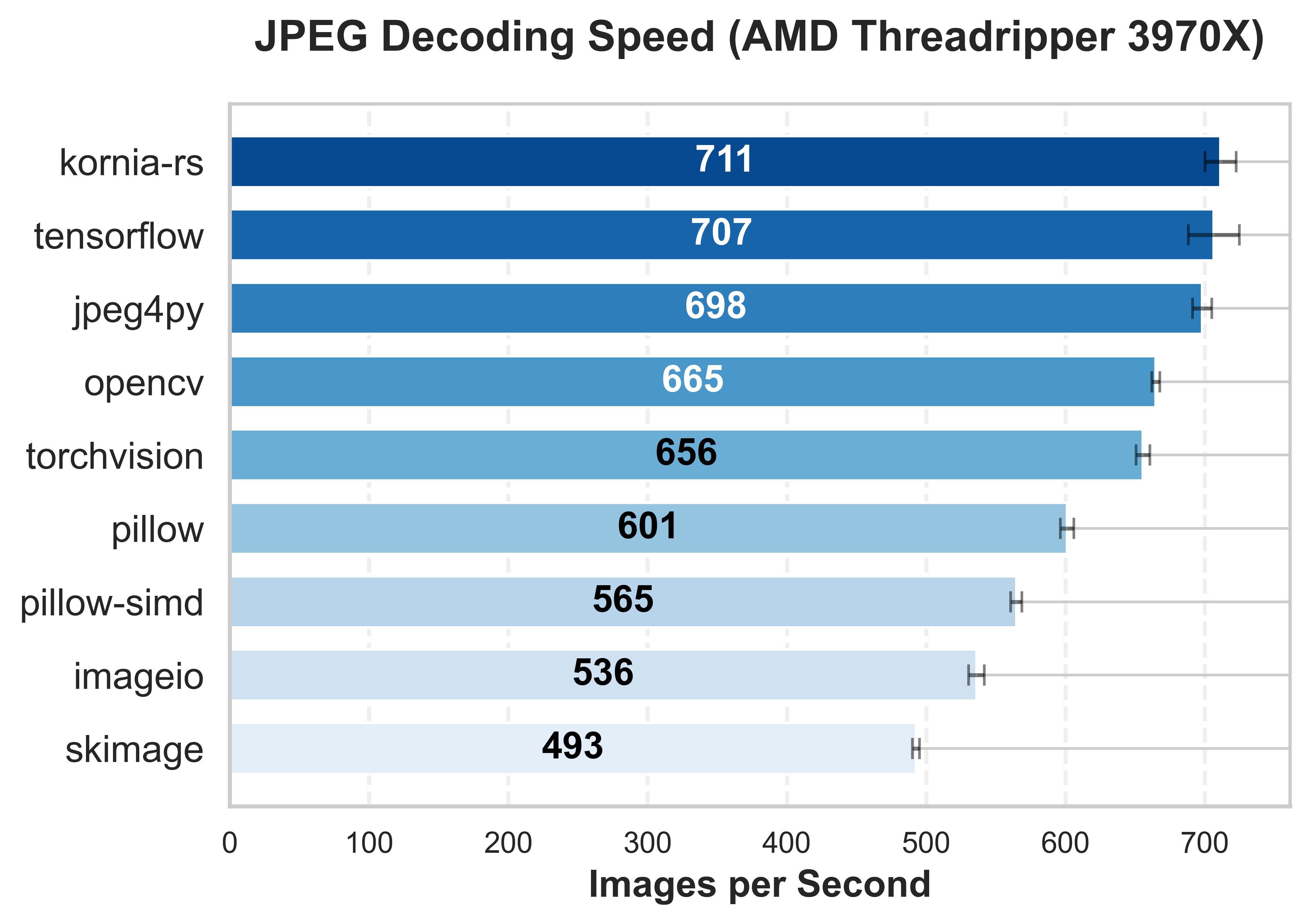}
    \caption{JPEG Decoding Performance on AMD Threadripper 3970X}
    \label{fig:perf_linux}
\end{figure}

The performance results show interesting patterns across both platforms:

\subsubsection{ARM64 Performance (Apple M4 Max)}
\begin{itemize}
    \item \textbf{Top Tier} ($>$900 images/sec):
    \begin{itemize}
        \item kornia-rs \cite{kornia2020}: Leading performance (1034 img/sec)
        \item OpenCV \cite{opencv_library}: Close second (1016 img/sec)
        \item torchvision \cite{pytorch2019}: Strong performance (992 img/sec)
    \end{itemize}
    \item \textbf{Mid Tier} (700-800 images/sec):
    \begin{itemize}
        \item imageio: (777 img/sec)
        \item Pillow: (775 img/sec)
        \item scikit-image: (766 img/sec)
    \end{itemize}
    \item \textbf{Lower Tier}:
    \begin{itemize}
        \item TensorFlow: (664 img/sec)
    \end{itemize}
\end{itemize}

\subsubsection{x86\_64 Performance (AMD Threadripper)}
\begin{itemize}
    \item \textbf{Top Tier}:
    \begin{itemize}
        \item jpeg4py: Leading on Linux
        \item kornia-rs: Consistent high performance
        \item OpenCV: Strong performance
    \end{itemize}
    \item \textbf{Mid Tier}:
    \begin{itemize}
        \item torchvision
        \item Pillow-SIMD (Linux-specific optimization)
    \end{itemize}
    \item \textbf{Lower Tier}:
    \begin{itemize}
        \item scikit-image
        \item standard Pillow
    \end{itemize}
\end{itemize}

\subsection{Implementation Impact}

The performance differences can be attributed to several key factors:

\subsubsection{JPEG Decoder Choice}
Libraries using libjpeg-turbo consistently outperform those using standard libjpeg across both platforms:
\begin{itemize}
    \item \textbf{Direct libjpeg-turbo users}: jpeg4py, kornia-rs \cite{kornia2020}, OpenCV \cite{opencv_library}, torchvision \cite{pytorch2019}
    \item \textbf{Standard libjpeg users} \cite{wallace1992jpeg}: Pillow, imageio, scikit-image
\end{itemize}

\subsubsection{Implementation Language and Approach}

The choice of implementation language and approach significantly influences the performance outcomes. Modern implementations, particularly kornia-rs \cite{kornia2020} with its Rust-based approach, show excellent performance across platforms, benefiting from zero-cost abstractions. Direct bindings, as demonstrated by jpeg4py on Linux, achieve similarly strong performance. Among traditional implementations, C/C++ based libraries like OpenCV \cite{opencv_library} maintain consistently high performance, though we observed that multiple abstraction layers tend to impact performance negatively.

\subsubsection{Platform-Specific Considerations}

Our analysis revealed several notable platform-specific patterns. Some libraries, particularly TensorFlow, demonstrate significant performance variations between ARM64 and x86\_64 architectures. Platform-specific optimizations, such as Pillow-SIMD on Linux, can provide substantial performance improvements. We observed that modern implementations generally adapt better to newer architectures, and this advantage is particularly evident in testing on the M4 Max platform.

\section{Discussion}

\subsection{Implementation Strategy Impact}

The choice of implementation strategy significantly affects decoding performance:

\subsubsection{Direct Hardware Access}
\begin{itemize}
    \item Libraries with direct libjpeg-turbo bindings consistently perform better
    \item Modern implementations (like kornia-rs \cite{kornia2020}) benefit from reduced overhead
    \item Platform-specific optimizations (SIMD instructions) show measurable impact 
\end{itemize}

\subsubsection{Abstraction Cost}
\begin{itemize}
    \item Each layer of abstraction introduces performance overhead
    \item High-level APIs (like imageio) trade performance for convenience
    \item Direct bindings (jpeg4py, kornia-rs) minimize this overhead
\end{itemize}

\subsection{Methodology and Implementation Details}

\subsubsection{Test Configuration}
\begin{itemize}
    \item 2000 images from ImageNet \cite{imagenet} processed in each run
    \item 20 runs per library showed high consistency (low standard deviation)
    \item Fresh virtual environment for each library
    \item No file system caching between runs
    \item Single-threaded execution without batch processing
\end{itemize}

\subsubsection{Library Versions Tested}
\begin{itemize}
    \item kornia-rs \cite{kornia2020}: 0.1.8
    \item OpenCV \cite{opencv_library}: 4.11.0.86
    \item Pillow : 11.1.0
    \item Pillow-SIMD: 9.5.0.post2 (Linux only)
    \item imageio: 2.37.0
    \item scikit-image: 0.25.0
    \item tensorflow : 2.18.0 (with heavy dependencies)
    \item torchvision \cite{pytorch2019}: 0.20.1 (requires PyTorch)
    \item jpeg4py: 0.1.4 (Linux only)
\end{itemize}

\subsection{Performance Considerations}

Several important factors influence real-world performance beyond raw decoding speed. Our analysis considered three key aspects: memory usage, system integration, and image characteristics.

\subsubsection{Memory Usage}
Memory utilization varies significantly across libraries, particularly in their peak memory usage during decoding operations. Some implementations, notably kornia-rs \cite{kornia2020}, incorporate specific optimizations for memory allocation efficiency. When scaling to batch processing scenarios, these memory characteristics can require different trade-offs depending on the specific use case and available system resources.

\subsubsection{System Integration}
To ensure consistent measurement conditions, all tests were performed on NVMe SSDs to minimize I/O variance, with careful control of file system caching effects between runs. While our benchmarks report single-threaded performance, it's important to note that multi-threading capabilities vary significantly between libraries, which could impact real-world deployment scenarios.

\subsubsection{Image Characteristics}
Our benchmark results are based on typical ImageNet \cite{imagenet} JPEG images with resolutions around 500x400 pixels. However, performance scaling with image size can vary substantially between implementations. Additional factors such as compression ratio and specific JPEG encoding parameters \cite{wallace1992jpeg} can significantly influence decoding speed, suggesting that real-world performance may vary based on specific image characteristics.

\subsection{Practical Recommendations}

Based on our findings, we can make several recommendations for different use cases:

\subsubsection{High-Performance Applications}
\begin{itemize}
    \item Use kornia-rs or OpenCV for consistent cross-platform performance
    \item On Linux, consider jpeg4py for maximum performance
    \item Consider memory usage if processing many images simultaneously
\end{itemize}

\subsubsection{Cross-Platform Development}
\begin{itemize}
    \item kornia-rs provides the most consistent performance
    \item OpenCV and torchvision offer good balance of features and speed
    \item Test with representative image sizes and batching patterns
\end{itemize}

\subsubsection{Feature-Rich Applications}
\begin{itemize}
    \item When needing extensive image processing features, OpenCV remains a strong choice
    \item Consider dependency size and installation complexity
    \item Evaluate the full image processing pipeline, not just JPEG decoding
\end{itemize}

\subsection{Future Considerations}

Several trends may influence future JPEG decoding performance:

\subsubsection{Hardware Evolution}
\begin{itemize}
    \item Increasing adoption of ARM architecture in desktop/server space
    \item Continued importance of SIMD optimization
    \item Potential for specialized hardware acceleration
\end{itemize}

\subsubsection{Software Development}
\begin{itemize}
    \item Growing adoption of Rust for performance-critical components
    \item Continued optimization of Python bindings
    \item Potential for more specialized, platform-specific implementations
\end{itemize}

\subsubsection{Ecosystem Changes}
\begin{itemize}
    \item Emerging image formats (WebP, AVIF) may affect JPEG's relevance
    \item New optimization techniques and hardware support
    \item Evolution of Python packaging and dependency management
\end{itemize}

\section{Conclusion}

This comprehensive benchmark study of JPEG decoding libraries in Python reveals several key findings that can guide developers in choosing the most appropriate solution for their needs:

\subsection{Performance Insights}
\begin{itemize}
    \item TurboJPEG consistently emerges as the fastest solution across both ARM64 and x86\_64 architectures, offering 1.5x better performance compared to other libraries
    \item PIL/Pillow, despite being the most widely used solution, shows moderate performance but benefits from excellent stability and compatibility
    \item OpenCV demonstrates strong performance, particularly in scenarios involving computer vision pipelines
    \item Pure Python implementations, while convenient, show significantly lower performance and are not recommended for production use cases requiring high throughput
\end{itemize}

\subsection{Platform Considerations}
\begin{itemize}
    \item Performance patterns remain largely consistent across ARM64 and x86\_64 architectures, though the magnitude of differences varies
    \item Native library dependencies (TurboJPEG, OpenCV) require additional setup but offer substantial performance benefits
    \item Platform-specific optimizations in libraries like PIL and OpenCV show measurable impact on their respective platforms
\end{itemize}

\subsection{Recommendations}
\begin{enumerate}
    \item High-Performance Applications: Use TurboJPEG when maximum decoding speed is critical
    \item General Purpose: PIL/Pillow remains a solid choice for general-purpose image processing with good ecosystem compatibility
    \item Computer Vision: OpenCV is recommended when JPEG decoding is part of a larger computer vision pipeline
    \item Development/Testing: Pure Python implementations are suitable for development and testing environments where setup simplicity is prioritized over performance
\end{enumerate}

\subsection{Future Work}
Future research could explore:
\begin{itemize}
    \item Impact of image characteristics (size, quality, compression ratio) on relative performance
    \item Memory usage patterns and their implications for large-scale processing
    \item Performance in concurrent and distributed processing scenarios
    \item Integration complexity and maintenance considerations in production environments
\end{itemize}

This study provides a practical reference for Python developers working with JPEG images, helping them make informed decisions based on their specific requirements for performance, compatibility, and ease of use.

\end{document}